\documentclass[twocolumn,english,prl,showpacs]{revtex4}
\usepackage[T1]{fontenc}
\usepackage[latin9]{inputenc}
\usepackage{amssymb}
\usepackage{graphicx}
\usepackage{amsmath,color}
\usepackage{mathrsfs}
\usepackage{float}
\usepackage{indentfirst}

\makeatletter

\def\journal #1, #2, #3, 1#4#5#6{{\sl #1~}{\bf #2}, #3 (1#4#5#6) }

\usepackage{mathrsfs}
\usepackage{float}
\usepackage{indentfirst}

\def\eqa{\begin{eqnarray}}
\def\eea{\end{eqnarray}}
\newcommand{\eq}{\begin{equation}}
\newcommand{\ee}{\end{equation}}

\makeatother

\begin{document}

\title{Charge-density-wave and topological transitions in interacting Haldane model}

\author{Lei Wang$^{1}$, Hao Shi$^{2}$, Shiwei Zhang$^{3}$, Xiaoqun Wang$^{2}$, Xi Dai$^1$ and X. C. Xie$^{4,5}$}

\affiliation{$^{1}$Beijing National Lab for Condensed Matter Physics and Institute of Physics, Chinese Academy of Sciences, Beijing 100190, China }

\affiliation{$^{2}$Department of Physics, Renmin University of China, Beijing 100872, China}

\affiliation{$^{3}$ Department of Physics, College of William and Mary, Williamsburg, Virginia 23187, USA}

\affiliation{$^{4}$ International Center for Quantum Materials, Peking University, Beijing 100871, China}

\affiliation{$^{5}$Department of Physics, Oklahoma State University, Stillwater, Oklahoma 74078, USA}

\begin{abstract}
Haldane model \cite{Haldane:1988p3868} is a noninteracting model for spinless fermions showing nontrivial topological properties. Effect of the electron-electron interaction on the topological phase pose an intriguing question. By means of the Hartree-Fock mean field, the exact diagonalization and the constrained-path Monte Carlo methods we mapped out the phase diagram of the interacting Haldane model. It is found that interaction breaks down the topological phase and drives the system into the charge-density-wave state. Sequence of the two transitions depends on the strength of next-nearest-neighbor hopping. Many-body Chern number and the charge excitation gap are used to characterize the topological transition.
\end{abstract}






\maketitle

\textit{Introduction}--Recently, there has been a raising interest on a class of topological phase of matter: the topological insulators, see \cite{Hasan:2010p23520, Moore:2010p15238} for reviews. Questions like what is the interaction effect on the topological insulators and how to characterize the topological phase beyond single particle basis pose intriguing challenges. \cite{Pesin:2010p15736,Rachel:2010p20458,Varney:2010p21916,Hohenadler:2010p24150,Zheng:2010p24293}. Haldane model \cite{Haldane:1988p3868} provides a prototype for investigating these questions. The model describes non-interacting spinless fermions on a honeycomb lattice. What makes it interesting is the complex next-nearest-neighbor hopping $t_{2}$, which opens a bulk gap and produces gapless edge states. The state is topologically nontrivial and is characterized by non-zero Chern number\cite{THOULESS:1982p24208}. The topological phase breaks down as staggered onsite energy $M$ exceeds a critical value.

The minimal form of interaction in Haldane model is the extended Hubbard interaction between the nearest-neighbors. The interacting Haldane model reads
\begin{eqnarray}
  H & = & H_0 + H_1 \nonumber \\
  H_0 & = &
  t_1\sum_{\langle i,j\rangle}c_i^{\dagger}c_j+it_2\sum_{\langle\langle i,j\rangle\rangle}\nu_{ij} c^{\dagger}_i c_j  \nonumber \\
  H_1 & = &V
  \sum_{\langle i,j \rangle} (n_i - \frac{1}{2}) (n_j - \frac{1}{2})
\label{eqn:Ham}
\end{eqnarray}
The nearest neighbor hopping $t_1$ is set as unit in the following. $\nu_{ij}=\pm1$ when the hopping path belongs to the (anti)clockwise loop.  The honeycomb lattice is a bipartite with two sites (denote as A and B) per unit cell. We focus on the half-filling case, \textit{i.e.} one particle per unit cell and  the system preserves the particle-hole symmetry. For vanishing $t_2$ the extended Hubbard interaction drives the system into a charge-density-wave (CDW) state at finite value of $V$\cite{Herbut:2006p1714, Raghu:2008p3865,Honerkamp:2008p4756} through a second order phase transition. For non-vanishing $t_2$, there are several questions.

\noindent 1. How does $t_2$ affect the CDW transition? What is the $t_2-V$ phase diagram looks like ?

\noindent 2. Presumably strong enough interaction would breaks the topological phase eventually. How does the topological transition organizes itself with the CDW transition? Are they separated or merged together?  There are several scenarios:
i) A continuous or weakly first order CDW transition occurs first and there is a finite region where small CDW order and the topological phase coexists.
ii) The two transitions merge: there is a single strongly first order transition where the CDW order develops and the topological phase breaks at the same time.
iii) The topological transition occurs first, resulting a topological trivial phase without CDW long range order.

\noindent 3. How to identify the topological transition for an interacting system ?  Unlike conventional phase transitions, the topological trivial/nontrivial phases do not correspond to different symmetry groups. The transition is of topological nature and can not be described by the conventional Ginzburg-Landau-Wilson theory. Moreover, the model is not exactly solvable once we introduce an interaction.


To address these questions, we perform  the Hartree-Fock (HF) mean-field, the exact diagonalization (ED) and the constrained-path quantum Monte Carlo (CPQMC) calculations. A key ingredient in our ED and CPQMC calculations is the twisted boundary conditions. Average over random twists greatly reduces the finite size effect\cite{POILBLANC:1991p20055,GROS:1992p20048,Gros:1996p20053,Lin:2001p19009}. Moreover, the twisted boundary conditions also provide a way of computing many-body Chern number \cite{Niu:1985p19667}: a direct characterization of topological orders. Since the topological transition is gap-closing by nature \cite{Haldane:1988p3868, Fukui:2005p19669,Thonhauser:2006p12423}, the charge excitation gaps are calculated as another indicator.




\begin{figure}[t!] \centering
    \includegraphics[height=6cm, width=8cm]{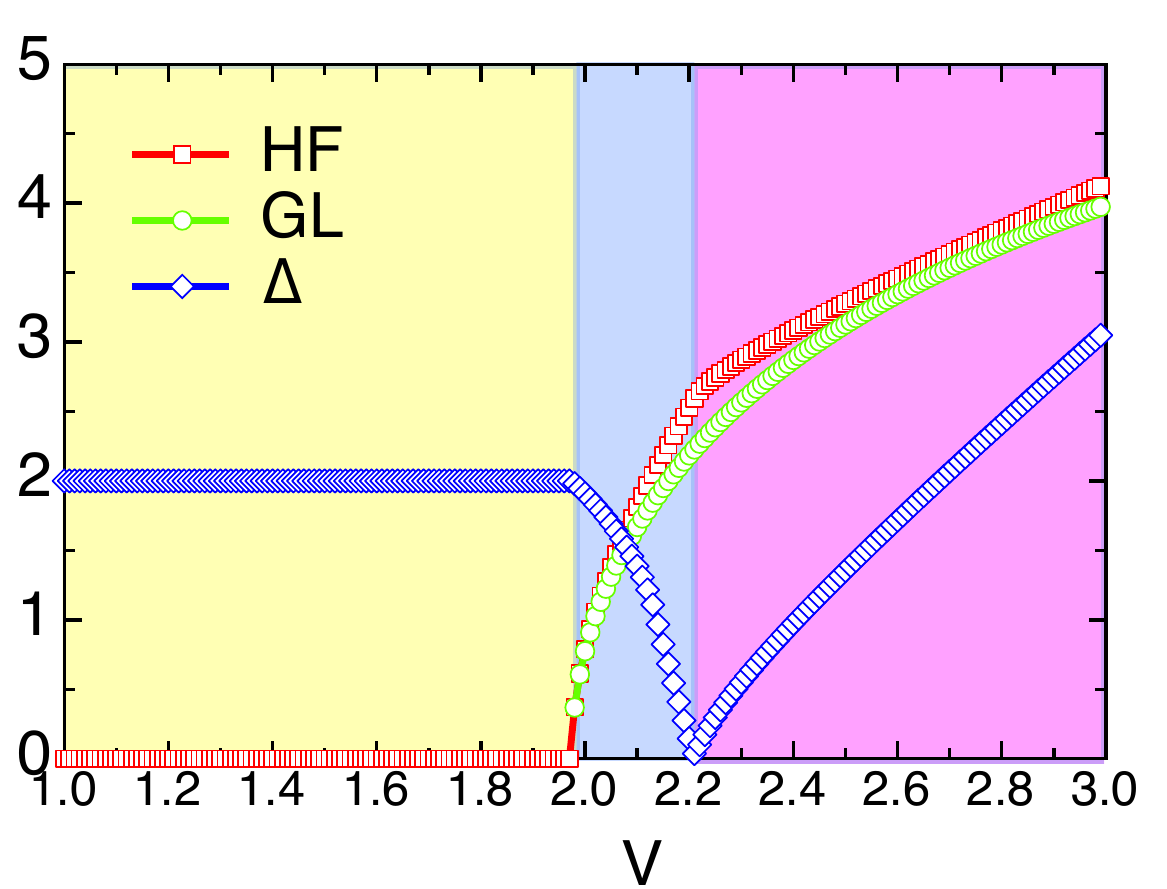}
    \caption{CDW order parameters as a function of $V$ calculated by HF (red squares) and GL (green dots) theory for $t_2=0.5$. Blue diamonds denote the charge excitation gap from HF calculations. } \label{fig:HF_GL_Delta}
\end{figure}

With these methods we map out the phase diagram of CDW and topological transitions and find that HF theory predict that a continuous CDW transition occurs before the topological phase transition. However many-body calculation reveals that the mean-field calculation misses to capture an important possibility that at small $t_{2}$ the topological phase transition happens without developing the CDW long range order.

\textit{Hartree-Fock mean-field and Ginzburg-Landau calculation}--Under the Hartree-Fock approximation the interaction term $Vn_in_j$ is approximated by $\langle n_i\rangle n_j+ n_i\langle n_j\rangle-\langle n_i\rangle \langle n_j\rangle$. The many-body Hamiltonian is then written into a summation of single particle terms. Introducing the CDW order parameter $M=zV(\langle n_{i\in A}\rangle-\langle n_{i\in B}\rangle)/2$ ( $z=3$ is the coordination number), then it couples to the density difference of two sub-lattices. The HF mean-field Hamiltonian is Fourier transformed into the momentum space and diagonalized on $200\times200$ meshes of the Brillouin zone (BZ). Converged CDW order parameter is determined self-consistently. 



\begin{figure}
    [tbp] \centering
    \includegraphics[height=6cm, width=8cm]{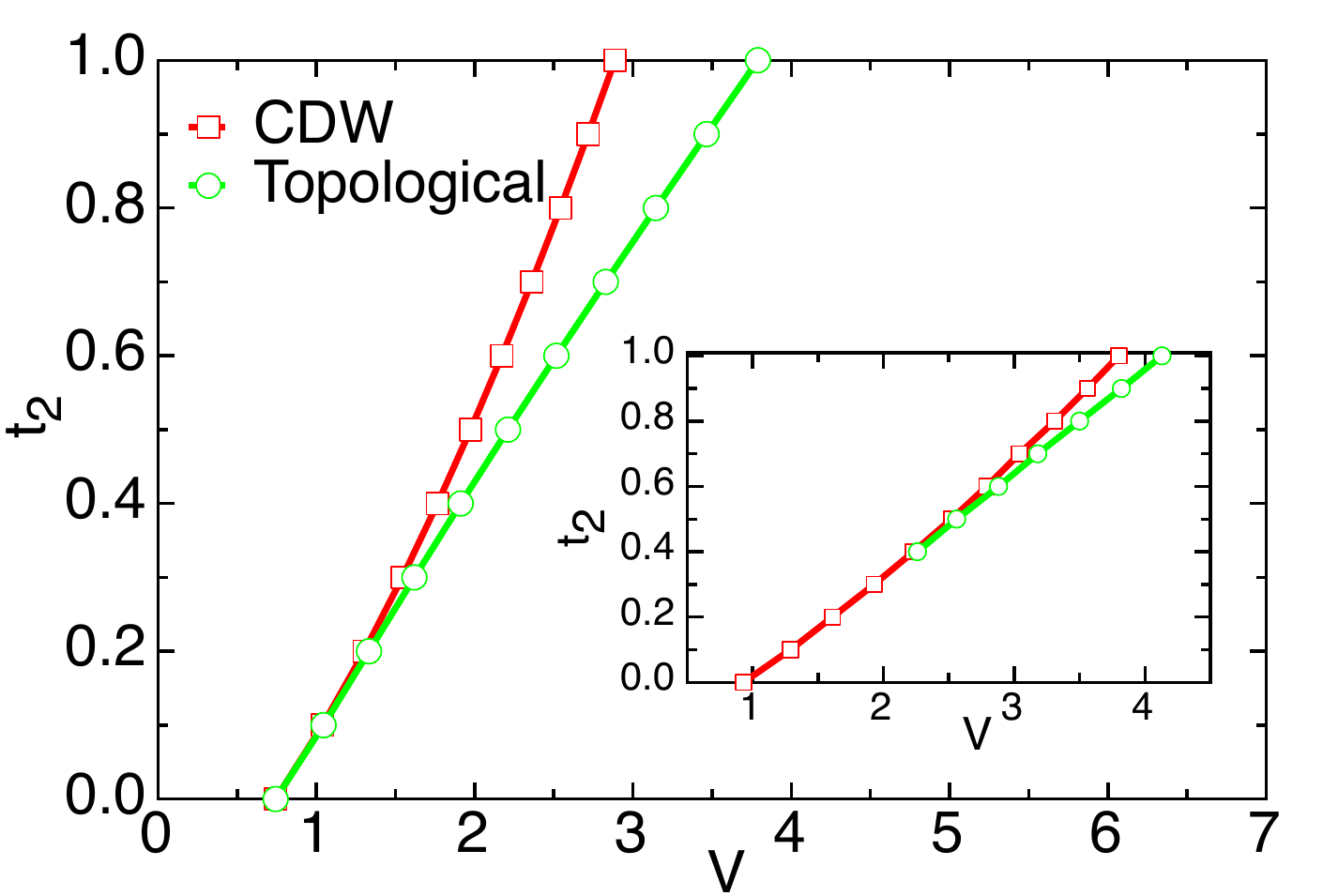}
	\caption{Mean-field $t_2-V$ phase diagrams. Red square line defines the CDW phase boundary. Green dot line denotes the topological transitions phase boundary, where the CDW order parameter exceeds $3\sqrt{3}t_2$. Inset shows the phase diagram taken into account of renormalization of $t_{1}$ (see text).}
		\label{fig:Phasediag}
\end{figure}

For $t_2=0$ we have a second order CDW transition at $zV=2.23$. Finite next-nearest-neighbor hopping $t_2$ pushes the transition point to a larger value of $V$. It is because $t_2$ term connects sites within the same sub-lattices and suppresses the CDW order. However, the CDW transition remains to be continuous. Interestingly, inside the CDW phase both the order parameter and the derivative of the ground state energy shows a kink (see Fig.\ref{fig:HF_GL_Delta}  recognized as the signature of the topological transition\cite{Cai:2008p20033}. Hartree-Fock theory reduces the interaction problem to the non-interacting Haldane model with a staggered onsite energy $M$. As $M$ exceeds $3\sqrt{3}t_2$, the topological phase transition occurs \cite{Haldane:1988p3868}. We also calculate the charge excitation gap $\Delta$ in the HF energy spectrum. It goes to zero right at the topological transition point, see Fig.\ref{fig:HF_GL_Delta}. The rising of the CDW order parameter and the closing of the charge excitation gap divide the phase diagram into three regions.
The size of the intermediate phase enlarges with the increasing of $t_2$ (see Fig.\ref{fig:Phasediag}), since  intuitively, the Haldane phase is more stable with a larger value of $t_2$.


To further reveal the nature of the transitions, we look into the Ginzburg-Landau theory of the model. The partition function reads $Z = \int \mathcal{D} c^{\dagger} \mathcal{D} c e^{- S}$, where the action is
$S [c^{\dagger}, c] = \int_0^{\beta} d \tau \psi_k^{\dagger} [(\partial_{\tau} - \mu)  + \vec{h}_k \cdot \vec{\sigma}]  \psi_k+ H_1$, $\vec{\sigma}$ are Pauli matrices. We have introduced $\psi_k = \left(\begin{array}{cc} c_{k A} & c_{k B}\end{array}\right)^T$ and write $H_0$ on this basis\cite{Haldane:1988p3868}. By performing the Hubbard-Stratonovich transformation: $e^{\int_0^{\beta} d \tau\frac{V}{2} \sum_{\langle i, j \rangle} (n_i - n_j)^2} = \int \mathcal{D} \phi e^{-\int_0^{\beta} d \tau \sum_{i \in A, \delta} V \phi_i (n_i - n_{i +\delta})+ \frac{V}{2} \phi_i^2}$ we introduce a bosonic field $\phi$ in each unit cell. It acts as the CDW order parameter, which couples to the difference of the densities on two sub-lattices. Integrating out fermions and keeping the zero frequency and momentum component (those contributions dominant over the instability) of bosonic field, we have





\begin{eqnarray}
   S & = & \frac{z V}{2} (1 - z V \chi)
  | \phi|^2 + \frac{(z V)^4}{4} b| \phi |^4
  \ldots
\label{eqn:Seff}
\end{eqnarray}
where $\chi=\sum_k\frac{{h_1}^2_k+{h_2}^2_k}{2|\vec{h}_k|}$
is the CDW susceptibility. Saddle point of the effective action of Eq.\ref{eqn:Seff} gives the Hartree-Fock theory. Second order CDW transition occurs at $1 - z V \chi = 0$. Near the transition point the order
parameter follows $z V \phi = \sqrt{\frac{|1 - z V \chi |}{b (z V)}}$.
In Fig.\ref{fig:HF_GL_Delta} we plot the GL prediction of the order parameter. The GL and HF result are in good accordance with each other near the CDW transition point. However, the kink feature of the CDW order parameter is missing in the GL result. It is because that GL expansion is only valid near the transition point and integration out of the fermions becomes invalid at the gap closing point.

We also like to remark that including the exchange terms $V\langle c_i^{\dagger}c_j\rangle c_{i}c_{j}^{\dagger}+h.c.$ in the mean field decoupling amounts to renormalize $t_1\rightarrow t_{1}+V\langle c_i^{\dagger}c_j\rangle$. It renders the CDW transition into a first-order transition for $t_{2}<0.4$ and there is no intermediate phase, see Fig.\ref{fig:Phasediag} Inset.




\begin{figure}[t]
   \centering
   \includegraphics[height=6cm, width=8cm]{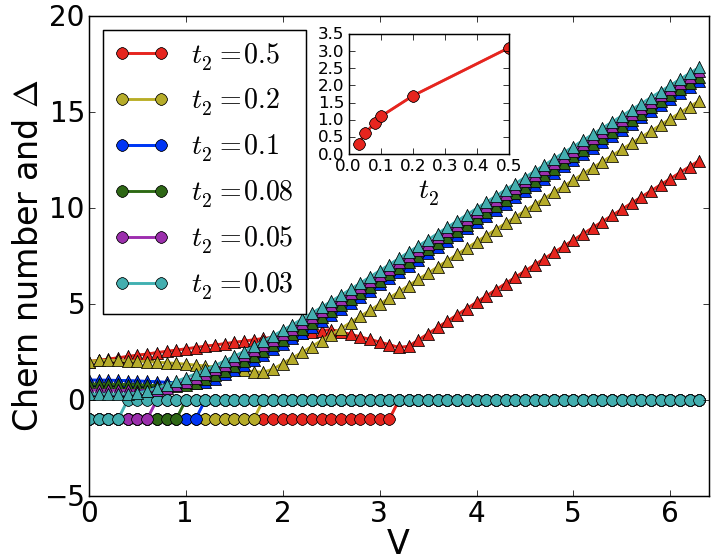}
   \caption{ED results of Chern number (dots) and charge excitation gap (triangles) of a $12$ sites lattice for different $t_{2}$. $20\times20$ twists are used to calculate the Chern number.  Inset shows the discontinuity of the Chern number point v.s. $t_{2}$.  }
   \label{fig:ED}
\end{figure}

\textit{Exact diagonalization and Constrained-path quantum Monte Carlo calculations}--In the following, we perform ED and CPQMC calculations\cite{Zhang:1995p13535,Zhang:1997p19631} of the model. In CPQMC, the many-body wave function is written as a summation of Slater determinants. Imaginary time evolution operator $e^{-\beta H}$ is applied on it to project out the ground state. For the present case, auxiliary Ising fields are introduced on bonds of the lattice to decouple $H_1$ into single-particle form. Summation over these Ising fields are performed by Monte Carlo sampling. Different with the spinful Kane-Mele-Hubbard model\cite{Hohenadler:2010p24150, Zheng:2010p24293}, QMC simulation encounters sign problems here. Moreover, due to the complex next-nearest neighbor hopping $t_2$ and the twisted-boundary-conditions, the wavefunctions and their overlaps become complex number. The phase problem is controlled approximately by the constrained path approximation where the overlaps between the Slater determinants and a given trial wave function are required to be positive\cite{Zhang:2003p13183}. Recently, the method has been used in the investigation of the spin density wave state \cite{Chang:2010p15661,Chang:2008p13530} and ferromagnetism \cite{Chang:2010p21383} in the Hubbard model.

Since the twisted boundary condition plays an important role in our calculations, we briefly explain it here. Under twisted boundary condition, the wave function gains a phase when electrons hop around lattice boundaries: $\Psi(...,\bold{r_j+L},...)=e^{i\bold{L}\cdot\bold{\Theta}}\Psi(...,\bold{r_j},...)$, where $\bold{\Theta}=(\theta_x,\theta_y)$. The Chern number is calculated by $C=\frac{i}{2\pi}\int\int d\theta_{x}d\theta_{y}[\langle \frac{\partial \Psi}{\partial \theta_{x}} | \frac{\partial \Psi}{\partial \theta_{y}}\rangle  -\langle \frac{\partial \Psi}{\partial \theta_{y}} | \frac{\partial \Psi}{\partial \theta_{x}} \rangle]$, where $|\Psi\rangle$ is the many-body ground state wavefunction with twist $\bold{\Theta}$ \cite{Niu:1985p19667}. Numerical implimentation of above equation follows \citet{Fukui:2005p19669} with one difference: the calculation was perform on many-body wavefunctions from ED, not single particle Bloch states. For other observables like energy and CDW structure factor, averaged over random twists $\langle O\rangle =\int\int d\theta_{x}d\theta_{y} \langle\Psi |O|\Psi\rangle$ greatly reduced the finite size  effect in ED and CPQMC calculations \cite{POILBLANC:1991p20055,GROS:1992p20048,Gros:1996p20053,Lin:2001p19009}.

\begin{figure}[!t] \centering
   \includegraphics[height=6cm, width=8cm]{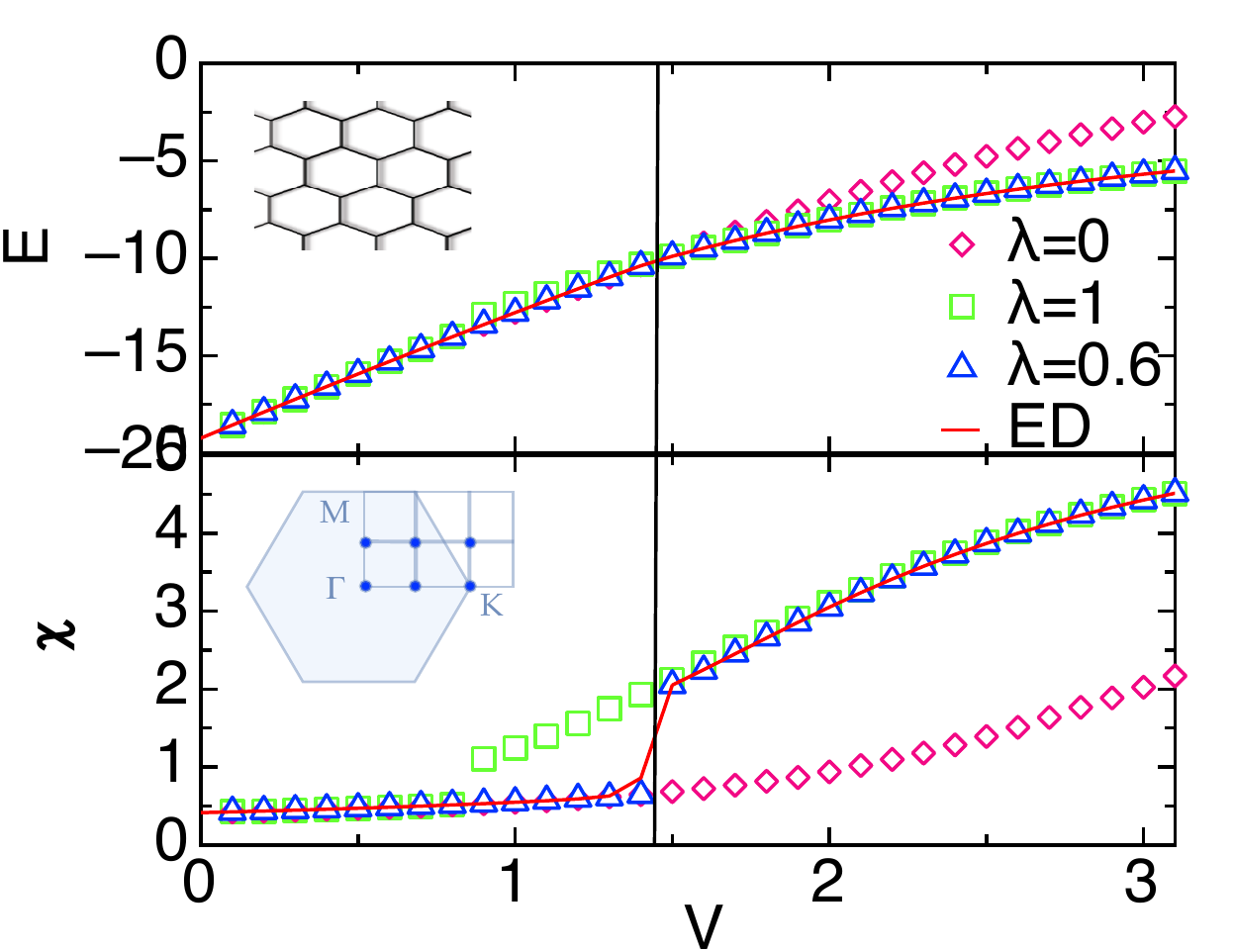}
    \caption{Comparison of the ED and MC results for a $24$ sites cluster with $t_2=0.1$. Inset shows the cluster geometry (PBC). Pink diamonds, green squares and blue triangles denote the MC results with $\lambda=0$ (noninteracting), $\lambda=1$ (Hartree-Fock) and  $\lambda=0.6$ reference states. Red line denotes ED results on the same cluster. } \label{fig:ED_MC}
\end{figure}

We first show Chern number versus interaction strength $V$ for different $t_2$ in Fig.\ref{fig:ED}. Even for very small cluster we see sharp transition of the Chern number (from $-1$ to $0$). We see that the topological transition point decreases continuously with $t_{2}$ (Inset of Fig.\ref{fig:ED}). On the other hand, it is known that the CDW transition point is finite even for vanishing $t_{2}$ \cite{Herbut:2006p1714, Raghu:2008p3865}. Therefore for small enough $t_{2}$ the topological transition would occur before the CDW transition. This result is not captured in the mean-field phase diagram Fig.\ref{fig:Phasediag} since there the only way of breaking the topological phase is to develop the CDW long range order. However in fact for small $t_2$ the Haldane phase is so fragile that short range CDW fluctuations may destroy it, resulting a gapped phase with neither topological nor CDW long range order. We also calculate the charge excitation gaps $\Delta=(E_{N+1}+E_{N-1}-2E_{N})/2$, where $E_N$ is the lowest energy in the half-filling sector, $E_{N\pm1}$ is the lowest energy in the sector with one more (less) particle. Since the particle-hole symmetry is respected, the above equation can be simplified to $\Delta=(E_{N-1}+zV/2-E_{N})$. From Fig.\ref{fig:ED} we see that the minimum point of the PBC gap is almost coincide with the jumping point of Chern number. Thus, we link the PBC gap-closing point with the topological transition. Since Chern number calculations with ED do not scale up to large clusters, in the following  we calculate the charge excitation gap with CPQMC for the signature of topological transition.

\begin{figure}[!t] \centering
    \includegraphics[height=6cm, width=8cm]{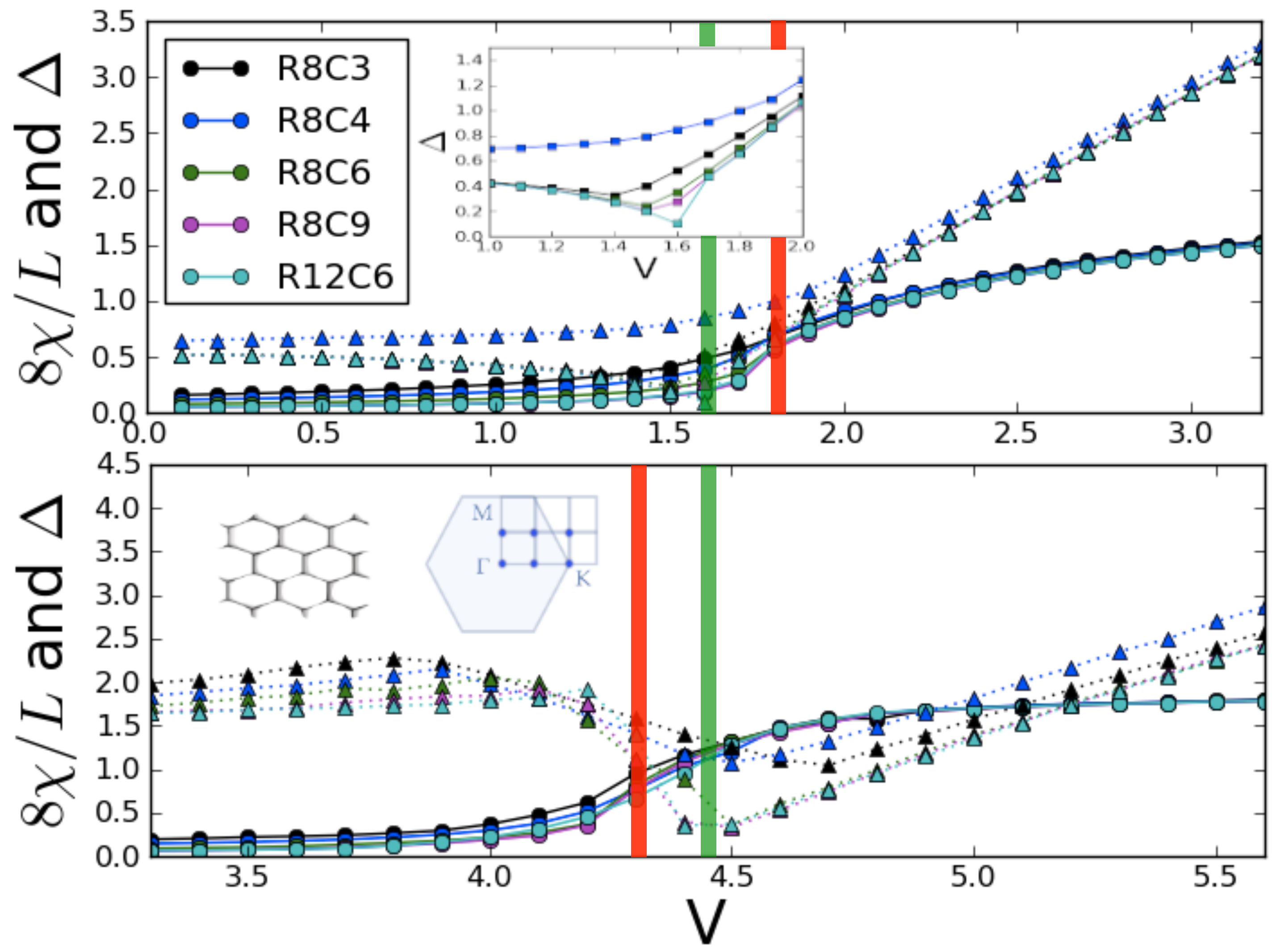}
    \caption{TABC CDW structure factor (dots) and PBC charge excitation gaps (triangles) vs. interaction strength $V$ for $t_2=0.1$ (upper panel) and $t_2=1.0$ (lower panel). Red and green line indicates the CDW and topological phase transition point from finite size scaling. Insets show a zoom in view of the charge excitation gap for $t_{2}=0.1$ and cluster geometries.} \label{fig:chi_Delta}
\end{figure}

Fig.\ref{fig:ED_MC} shows the benchmarks of the CPQMC results with the ED results. If PBC is adopted and the number of columns is a multiply of $3$ then K-point is a validate point in the momentum space. The reference state for CPQMC calculation is obtained from the HF calculation with $V^\prime=\lambda V$ where $\lambda$ is an interpolation parameter, resembles the tuning parameter in hybrid functionals. The noninteracting state corresponds to $\lambda=0$ and the "true HF state" corresponds to $\lambda=1$. We sample the ground state energy and the CDW structure factor $\chi^{CDW} = \frac{1}{N}\sum_{i,j}\langle (n_i^A - n_i^B) (n_j^A - n_j^B)\rangle$ in MC calculations. The two MC energies crosse at $V=3.0$. On both sides, we take the physical quantities from the one with lower energies. Practically, we also adopt the $\lambda=0.8$ for $t_2=1.0$ and $\lambda=0.6$ for $t_2=0.1$. Both approaches give results show good accordance with exact ones, see Fig.\ref{fig:ED_MC}. For the $24$ sites lattice, the CDW structure factor shows a jump at $V=3.0$. 




 \begin{figure}[t]
   \centering
   \includegraphics[height=6cm, width=8cm]{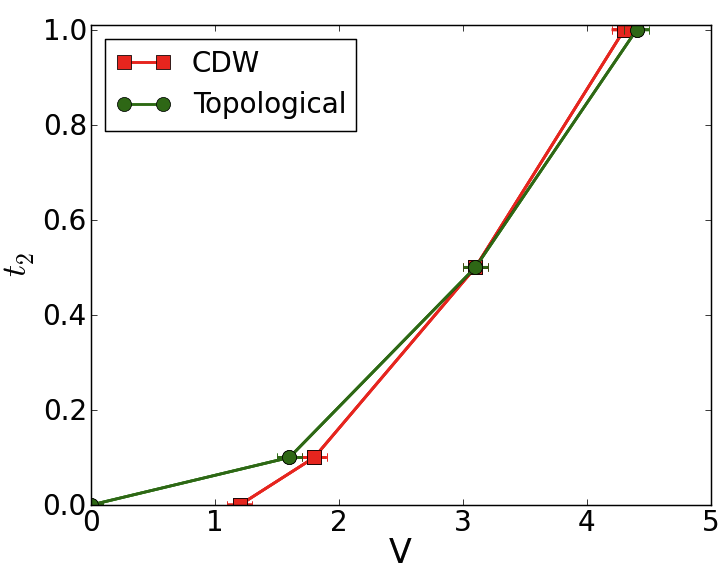}
   \caption{Phase diagram determined by CPQMC calculations. Red squares and green dots denote the phase boundaries of the CDW and topological transitions. Error bars indicates the uncertainties of determining the phase transition boundaries. Lines are guides to the eye.}
   \label{fig:many_body_phasediag}
\end{figure}

We show the CDW structure factor and the charge excitation gap from CPQMC calculations in Fig.\ref{fig:chi_Delta}. The CDW structure factors are continuous and obey the scaling law of second order transition $\chi L^ {b}= f [(V - V_c) L^{a}]$, with $f$ a universal function independent of system size. By fitting the scaling law we extract the CDW transition point $V_{c}$. Gap closing is interpreted as signature of the topological transition. We mark the CDW and topological transition point in the figure. Widths of them  indicates uncertainty of the fitting. Even with the uncertainties, we could clearly see the topological transition occurs before the CDW transition for $t_{2}=0.1$ and after it for $t_{2}=1.0$.




We conclude our many-body calculations with phase diagram shown in Fig.\ref{fig:many_body_phasediag}. Different with the mean-field prediction in Fig.\ref{fig:Phasediag}, the CDW and topological transition lines intersect. Above the intersection, the nature of two transitions is mean-field like, \textit{i.e.} long-range CDW order develops and then destroys the Haldane phase. Albeit the two phase boundaries shift against the mean-field values. Below the intersection the topological phase is destroyed before the CDW order develops. As $t_{2}\rightarrow0$ the topological boundary goes to zero while the CDW transition point remains at a finite value.

\textit{Outlook}--First, to better characterize the topological transition in an interacting system, we will look into the calculation of many-body Chern number or entanglement entropy/spectrum within CPQMC method. Second, the nature of topological transition in small $t_{2}$ region deserves further study. Is the topological phase been destroyed purely due to fluctuation effect or instability toward other order, say Kekule distortion? Third, results reported here may shed light on phase diagram of the Kane-Mele-Hubbard model where Z$_{2}$ topological order competes with antiferromagnetic order.


\textit{Acknowledgment}--LW thanks Yuan Wan and Zi Cai for discussions. Numerical calculations were performed on the Kohn cluster of RUC and the Dawning cluster of IOP. The work is supported by NSF-China and MOST-China. XCX is also supported in part by DOE through DE-FG02-04ER46124.

\bibliography{/Users/leiwang/Documents/Papers/papers}


\end{document}